\documentclass[aps,pre,superscriptaddress,floats,showpacs,floatfix,showkeys,preprint]{revtex4-1}

\usepackage{bm}
\usepackage{graphicx}
\usepackage{subfigure}
\usepackage{amsmath}
\usepackage{amssymb}
\usepackage{color}
\usepackage[a4paper,left=1.5cm, right=1.3cm, top=3.0cm,bottom=2cm]{geometry}

\newcommand\Rey{\mbox{\textit{Re}}}  
\newcommand{\Pra}{\mbox{\textit{Pr}}} 
\newcommand{\Ray}{\mbox{\textit{Ra}}}
\newcommand{\Tay}{\mbox{\textit{Ta}}}
\newcommand{\Nus}{\mbox{\textit{Nu}}}
\newcommand{\e}[1]{\ensuremath{\times 10^{#1}}}
\newcommand\avg[1]{\ensuremath{\langle#1\rangle}}

\begin{document}

\title{Heat transport in Rayleigh--B\'enard convection and angular momentum transport in Taylor--Couette flow: a comparative study}
\author{Hannes Brauckmann}
\affiliation{Fachbereich Physik, Philipps-Universit\"at Marburg, D-35032 Marburg, Germany}
\author{Bruno Eckhardt}
\affiliation{Fachbereich Physik, Philipps-Universit\"at Marburg, D-35032 Marburg, Germany}
\affiliation{J. M. Burgerscentrum, Delft University of Technology, 2628 CD Delft, The Netherlands}
\author{J\"org Schumacher}
\affiliation{Institut f\"ur Thermo- und Fluiddynamik, Technische Universit\"at Ilmenau, Postfach 100565, D-98684 Ilmenau, Germany}
\date{\today}

\begin{abstract}
Rayleigh--B\'{e}nard convection and Taylor--Couette flow are two canonical flows that have many properties in common.
We here compare the two flows in detail for parameter values where the Nusselt numbers, i.e. the 
thermal transport and the angular momentum transport normalized by the corresponding laminar values,
coincide. 
We study turbulent Rayleigh--B\'{e}nard convection in air at Rayleigh number $\Ray=10^7$ and 
Taylor--Couette flow at shear Reynolds number  $\Rey_S=2\times 10^4$ for two different mean rotation rates
but the same Nusselt numbers. For individual pairwise related fields and convective currents, we compare the 
probability density functions normalized by the corresponding root mean square values and taken at different 
distances from the wall. We find one rotation number for which there is
very good agreement between the mean profiles of the two corresponding quantities
temperature and angular momentum. Similarly, there is good agreement between
the fluctuations in temperature and velocity components. For the heat and angular momentum currents, there are differences 
in the fluctuations outside the boundary layers that increase with overall rotation
and can be related to differences 
in the flow structures in the boundary layer and in the bulk. The study extends the similarities between
the two flows from global quantities to local quantities and reveals the effects of rotation on the transport.
\end{abstract}

\maketitle

\section{Introduction}
Convection in layers of fluids heated from below and cooled from above (Rayleigh--B\'enard or RB flow) 
and the flow between two rotating cylinders (Taylor--Couette or TC flow) are among the canonical flows in
fluid mechanics. 
Studies of their stability properties and the manner in which the laminar profiles give way to more structured and complicated 
flows have provided much insight into the transition to turbulence with linear  instabilities  \cite{Chandrasekhar1961,Koschmieder1993}.  
The behavior well above the onset of turbulence has also been investigated starting with the experiments by Wendt \cite{Wendt1933}. 
Many different flow regimes that are not yet fully explained or explored have been described \cite{Ostilla-Monico2014b,Grossmann2016}. It was 
realized early on that despite the differences in the driving forces, there are many similarities, and it is helpful to  draw analogies 
and to compare the properties of both flows \cite{Low1925}. The intimate relations between the two flows have led Busse 
\cite{Busse2012} to characterize them as the \textit{twins of turbulence}.

A formal analogy between RB and TC flow (and pipe flow as well) was developed and described in Eckhardt et al. 
\cite{Eckhardt2007,Eckhardt2007a} (see also Bradshaw \cite{Bradshaw1969} for an earlier approximate relation 
and Dubrulle \& Hersant \cite{Dubrulle2002} for a similar analogy). 
The analogy identifies pairs of equations that describe the total energy dissipation and the global transport of heat or angular momentum, 
respectively, in the two flows. The equations allow one to relate transport properties, dimensionless parameters and other quantities,
and have been used in particular to study scaling relations in fully developed turbulent flows \cite{Grossmann2016}. 
The similarity in the equations suggests that a more detailed  comparison between the two flows should be possible. 

We here explore this option within direct numerical simulations (DNS). We describe the difficulties one has to overcome in 
identifying corresponding parameters, and present case studies where
detailed comparisons are possible. In particular, we will compare the turbulent transport currents with respect to their 
statistical properties. Furthermore, we can relate components of the involved turbulent fields to each other and compare
their statistical fluctuations at different distances from the walls. The focus of our study is on the general ideas and an illustration
for a few examples, but not on a comprehensive study for all parameter values. 
Specifically, we will take one set of data for RB flow and compare it to TC flow cases 
at two different rotation numbers, which allows us to study the effect of rotation.
The data are taken from well-resolved DNS of both flows at moderate Rayleigh and 
Reynolds numbers.

The outline of the manuscript is as follows. In section~\ref{sect:relations} we present
the balance equations, the numerical methods and discuss the analogy. 
In section~\ref{sect:reference} the choice of corresponding parameters for the
comparison is explained. In section~\ref{sect:statistics} 
the area-averaged mean currents and their probability density functions (PDFs) as well as other
pairwise related properties at different distances from the wall are analyzed. 
We conclude the
work with a short discussion of the particular structures of the convective currents and 
a summary in section~\ref{sect:conclusions}. 

\section{Relations for transport currents and dissipation rates}
\label{sect:relations}

RB flow is modeled by the three-dimensional Boussinesq equations for the velocity field ${\bf u}$ and the
temperature field $T$ \cite{Ahlers2009, Chilla2012}. The equations are solved using the Nek5000 software \cite{nek5000}, 
a spectral element method \cite{Fischer1997, Scheel2013}. 
The physical system is characterized by the imposed temperature difference between bottom and 
top plates, $\Delta$, the height $d$ of the domain, and the free-fall velocity $U_f=\sqrt{g\alpha\Delta d}$ with the thermal expansion 
coefficient $\alpha$ and the acceleration due to gravity, $g$. The kinematic viscosity $\nu$ of the fluid and the thermal diffusivity 
$\kappa$ are combined in the Prandtl number $\Pra=\nu/\kappa$. 
The flow is confined to a cylinder with insulating sidewalls.

The mean heat flux across the layer, i.e., in the $z$-direction, is given by
\begin{equation}
J_T=\avg{ u_z T} - 
\kappa \frac{\partial \avg{ T }}{\partial z} = 
\Nus_T\, J_{T , lam} \,,
\label{J_temp}
\end{equation}
with $J_{T, lam}$ the purely diffusive heat flux below the onset of convective motion,
\begin{equation}
J_{T, lam}=\kappa \frac{\Delta}{d}\,.
\label{J_temp1}
\end{equation}
Here, $\avg{\cdot}=\langle\cdot\rangle_{A,t}$ denotes an 
average over horizontal planes at fixed height and over time. 
Equation (\ref{J_temp}) already contains the definition of the Nusselt number $\Nus_T$  
which measures the heat transport relative to the laminar situation. A second relevant equation is that for the mean kinetic 
energy dissipation rate of the velocity field $\epsilon_u$. It is obtained by multiplying the momentum balance of the 
Navier--Stokes--Boussinesq equations with 
the velocity $\bf{u}$ and integrating over the volume and over time, 
\begin{equation}
\epsilon_u = \frac{\nu^3}{d^4} \Pra^{-2} \Ray (\Nus_T-1) \ .
\label{eps_temp}
\end{equation}
Here, $\Ray = \alpha g d^3 \Delta / (\kappa \nu)$ is the Rayleigh number, the second dimensionless parameter. 
The data set which we  
use for the comparisons is obtained in a closed cylindrical cell with a unity aspect ratio (diameter=height)
at $\Ray=10^7$ and $\Pra=0.7$. 

The TC system is characterised by the radii $r_1$ and $r_2$ of the inner and outer cylinder, 
which rotate with the angular velocities $\omega_1$ and $\omega_2$, respectively.
The flow between the cylinders is governed by the incompressible Navier--Stokes equations 
for the velocity ${\bf u}=(u_r, u_\varphi, u_z)$ in cylindrical coordinates $(r,\varphi,z)$.
We solve the equations with periodic boundary conditions in the axial direction using
a spectral method \cite{Meseguer2007}. 
In TC flow, the gap width $d=r_2-r_1$ and the velocity difference between the cylinders 
$U_0=2r_1 r_2(r_1+r_2)^{-1}(\omega_1-\omega_2)$  (calculated in a rotating frame of 
reference \cite{Dubrulle2005,Brauckmann2016a})  serve as characteristic scales for lengths and 
velocities. We choose a system height of $2d$ 
so that one pair of Taylor vortices fits into the computational domain. 
Thus, the diameter of a single Taylor vortex is similar to that of the large-scale circulation in RB flow with aspect ratio one.
Further details of the simulation procedure are discussed in references \cite{Brauckmann2013,Brauckmann2016a}.

For the derivation of expressions in TC flow that correspond to (\ref{J_temp}) and (\ref{eps_temp}) in RB
flow, 
we start with the azimuthal velocity $u_\varphi$.
Averaging the $\varphi$-component of the Navier--Stokes equation over time and over cylinders at fixed radii $r$  between the positions 
of the inner ($r_1$) and outer ($r_2$) cylinder, one finds  \cite{Eckhardt2007,Eckhardt2007a}
\begin{equation}
J_{\omega} = r^3\left( \avg{ u_r \omega} 
- \nu\frac{\partial\avg{\omega}}{\partial r} \right) = \Nus_\omega J_{\omega,lam}\,,
\label{J-omega}
\end{equation}
with the angular velocity $\omega=u_\varphi/r$ 
and $J_{\omega,lam}$ the angular momentum flux in the laminar case,
\begin{equation}
J_{\omega,lam} = \nu \frac{r_1^2 r_2^2}{r_a d} (\omega_1 - \omega_2) \,.
\label{J-omega-lam}
\end{equation}
Here, $r_a=(r_1+r_2)/2$ is the mean radius. The averaged current $J_{\omega}$ is independent of the radius and 
conserved in time. Physically, it corresponds to the torque needed to keep the cylinders
in motion; it corresponds naturally to the heat transport (\ref{J_temp}) in RB flow,
which is why we also introduced a Nusselt number $\Nus_\omega$ corresponding to $\Nus_T$.
Similarly, one can multiply the Navier--Stokes equation with the velocity ${\bf u}$ and average over
volume and time to obtain the mean kinetic energy dissipation rate, corresponding to (\ref{eps_temp}). However, the dissipation
associated with the laminar profile has to be taken out, so that we are led to consider
\begin{equation}
\epsilon_u=\epsilon-\epsilon_{lam} = \frac{(\omega_1 - \omega_2)}{r_a d} J_{\omega,lam} (\Nus_\omega - 1) 
= \frac{\nu^3}{d^4} \sigma^{-2} \Tay(\Nus_{\omega}-1)\,.
\label{eps_omega}
\end{equation}
with the geometric parameter $\sigma=r_a^4/(r_1 r_2)^2$ denoted as quasi-Prandtl number.  
The dimensionless Taylor number is defined as $\Tay=\sigma d^2 r_a^2 
(\omega_1-\omega_2)^2/\nu^2$. Furthermore, the radius ratio is denoted by $\eta=r_1/r_2$ and the specific 
angular momentum is defined by ${\mathcal L}=r u_{\varphi}$.

To separate the influences of shear and rotation, 
we adopt the parameters introduced by Dubrulle et al. \cite{Dubrulle2005}. 
With the Reynolds numbers 
\begin{equation}
\Rey_1=\frac{r_1\omega_1 d}{\nu} \quad\mbox{and}\quad \Rey_2=\frac{r_2\omega_2 d}{\nu}\,.
\end{equation}
for the inner and outer cylinders, respectively, we form 
the shear Reynolds number and the rotation number  
\begin{equation}
\Rey_S=\frac{2}{1+\eta} (\Rey_1-\eta \Rey_2)\quad\mbox{and}\quad R_\Omega=(1-\eta)\frac{\Rey_1+\Rey_2}{\Rey_1-\eta \Rey_2}\,.
\end{equation}
The relation to the Taylor number is given by $\Tay=\sigma^2\Rey_S^2$.

A comparison between (\ref{eps_temp}) and (\ref{eps_omega}) suggests an association $\Pra\equiv\sigma$ between
the Prandtl number $\Pra$ and the quasi-Prandtl number $\sigma$, and $\Ray\equiv\Tay$ between the 
Rayleigh number $\Ray$ and the Taylor number $\Tay$. However, there are various reasons why this 
is not sufficient. For example, a direct comparison between (\ref{eps_temp}) and (\ref{eps_omega}) 
suggests equality of the combinations
$\Pra^{-2}\Ray$ and $\sigma^{-2}\Tay$, only and does not relate $\Ray$ and $\Tay$ directly.
Moreover, TC flow has two Reynolds numbers, and the Taylor number captures 
only their difference. The overall rotation, 
as measured by the rotation number $R_\Omega$, does not enter, but it is 
known that the torque varies non-monotonically with $R_\Omega$ \cite{Paoletti2011,VanGils2011,Brauckmann2013,Ostilla2013,Merbold2013}. 
Similarly, critical values for the onset of instability are 
given by $\Ray_c=1708$ \cite{Chandrasekhar1961} for RB flow and by
$\Tay_c=1708/[R_\Omega(1-R_\Omega)]$ \cite{Dubrulle2005} for TC flow
(in the limit $\eta\rightarrow1$ where $\Tay=\Rey_S^2$), again 
highlighting the significance of the rotation number. We therefore have to look for alternatives on how to relate the two flows. 

\section{Choice of reference point for comparison}
\label{sect:reference}
\begin{figure}
  \centerline{
  	\includegraphics[width=\textwidth]{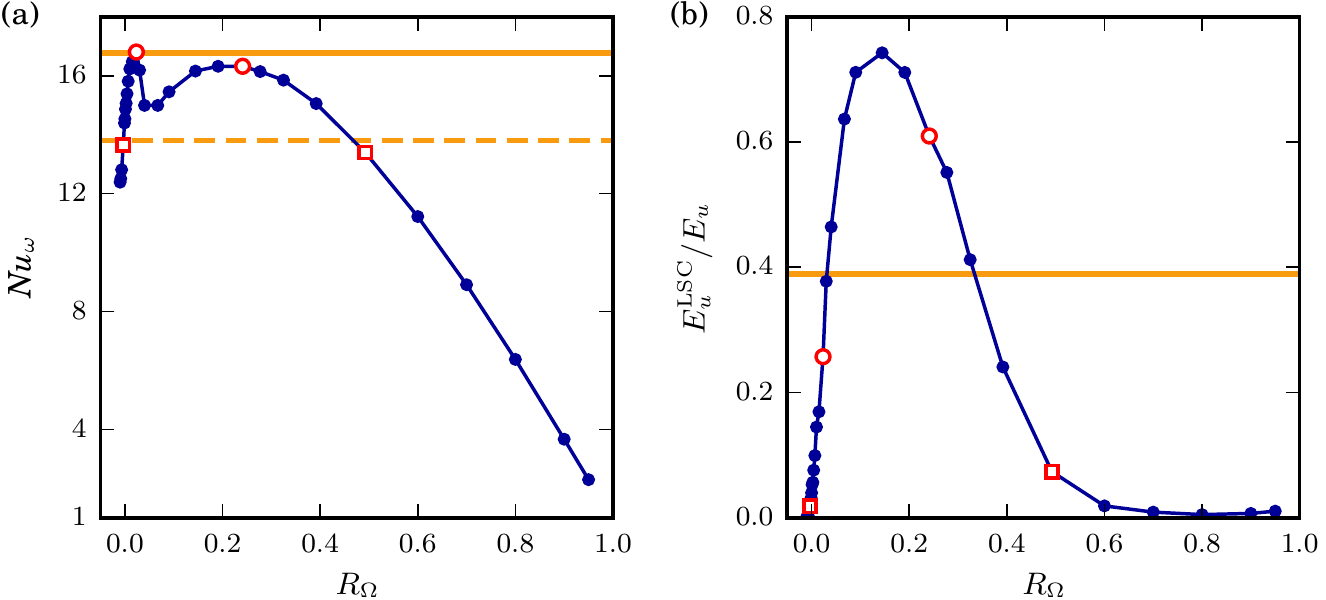}} 
\caption{(a) Variation of the Nusselt number with the system rotation $R_\Omega$ for TC flow with $\eta=0.99$ and $\Rey_S=2\e{4}$ (circles). 
(b) Ratio of the energy contained in the large-scale circulation $E_{u}^{LSC}$ to the energy 
of the total cross-flow $E_u$ as given in eq. (\ref{ratioEkin}).  
The horizontal lines in (a,b) indicate the corresponding values from 
RB convection with $\Pra=0.7$ and $\Ray=10^7$ (solid) and $\Ray=5\e{6}$~(dashed). The four possible TC data points for the comparison 
(with $\Nus_\omega$ close to a RB Nusselt number) are marked as open symbols.}
\label{fig:Nu-LSC}
\end{figure}
In the following, we discuss how the reference point for the one-to-one comparison is chosen. 
As described above, equating $\Ray$ and $\Tay$ directly is not possible because of 
ambiguities in their definitions. 
A meaningful comparison can be based on the Nusselt number, because it defines the boundary
layer thickness and hence the mean profiles. Similarly, the Reynolds stresses, when normalized by $\Nus J_{lam}$, 
should fluctuate with mean value 1 in regions where the viscous contributions to the transport are small. 
This allows for an absolute comparison of probability density functions 
since the dimensionless version with $\Nus$ scaled out has the same mean 
and, as we will show for one of the cases here, also the same variance.

We also have to select the curvature parameter $\eta$ in TC flow, which has no counterpart in RB flow 
because the heated and cooled plates are planar. We therefore take $\eta=0.99$ since the 
curvature effects disappear for $\eta$-values close to $1$. Finally, the mean 
system rotation in TC flow which is defined by $R_\Omega$ has to be selected for the direct comparison. 
Again, an analogous parameter is missing in RB convection. 
In figure~\ref{fig:Nu-LSC}(a) a curve $\Nus_{\omega}(R_\Omega)$ at $\eta=0.99$ is shown for 
$\Rey_S=2\times 10^4$.  
A Nusselt number $\Nus_{\omega}$ that is comparable to the RB flow value of $\Nus_T=16.7$ at $\Ray=10^7$ 
(solid horizontal  line) was obtained 
for $R_{\Omega}=0.023$ and $R_{\Omega}=0.241$ (open circles). 
These two runs will be denoted as case~1 and case~2, respectively,
and will be used to study the effects of the rotation number. 
In both cases the cylinders are co-rotating with angular velocity ratios $\mu=\omega_2/\omega_1=0.40$ and $\mu=0.92$ for $R_\Omega=0.023$ and $R_\Omega=0.241$, respectively.
Decreasing the Rayleigh number gives other crossing points, as indicated by the dashed
line and the open squares in figure~\ref{fig:Nu-LSC}(a) for $\Ray=5\e 6$.
For this lower $\Ray$ the RB and TC flows differ noticeably, so that we will subsequently focus on the 
cases 1 and 2 only.
We furthermore note that in the first case the relative distance to the linear instability 
$\Tay/\Tay_c\approx5.35\e{3}$ is close to the corresponding RB value $\Ray/\Ray_c\approx5.86\e{3}$ for $\Ray=10^7$, 
whereas in the second case the ratio $\Tay/\Tay_c\approx4.29\e{4}$ is much higher.

Figure~\ref{fig:Nu-LSC}(b) shows the ratio    
\begin{equation}
\frac{E^{LSC}_u}{E_u}=\frac{\avg{\avg{u_r}_{\varphi,t}^2+\avg{u_z}_{\varphi,t}^2}_{r,z}}{\avg{u_r^2+u_z^2}_{V,t}}\,,
\label{ratioEkin}
\end{equation}
of the energy contained in the mean vortical motion to the energy of the total cross-flow. It measures the relative 
strengths of temporally and streamwise-averaged Taylor vortices, which are analogous to the large-scale circulation 
in RB flow. The vortex strength varies with rotation, and the curve shows that case 2 is more strongly dominated by 
the large-scale vortices than case 1. In RB flow with $\Ray=10^7$, the corresponding energy ratio of approximately 
$0.4$ is of similar magnitude and lies between the two TC cases.

\begin{figure}
\begin{center}
              \includegraphics{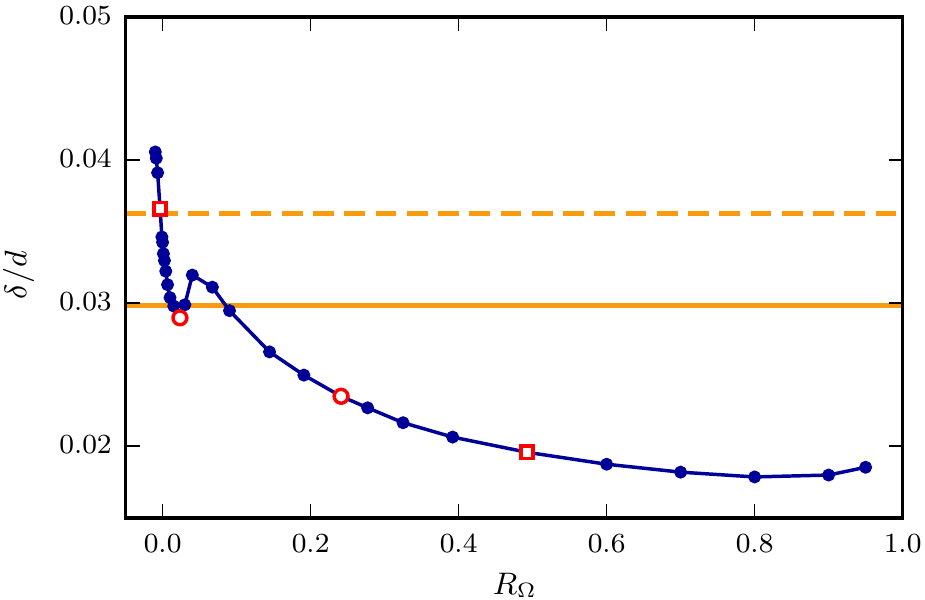}
\caption{The boundary layer thickness $\delta_{\mathcal L}$ which is based on the slope of the angular momentum profile at the wall is displayed 
versus $R_{\Omega}$. The corresponding thermal boundary layer thickness $\delta_T=d/(2 \Nus_T)$ of the RB flow is indicated 
by the horizontal lines. Parameter values are $\eta=0.99$ with $\Rey_S=2\e{4}$ for TC flow and $\Pra=0.7$ with $\Ray=5\e{6}$ 
(dashed) and $\Ray=10^7$ (solid) for RB convection. The four possible TC data points for the comparison are again marked
as open symbols.}
\label{fig:slope-BL}
\end{center}
\end{figure}

As additional measures for the comparison of both flows, we analyse the boundary layer thicknesses.
In analogy to the thermal boundary layer thickness 
$\delta_T=-\Delta/(2\partial_z\avg{T}|_{z=0,d})=d/(2\Nus_T)$ in RB flow,
we define the boundary layer thicknesses at the inner cylinder ($r=r_1$)
\begin{equation}
 \delta_\omega=\frac{\Delta_\omega}{-2\partial_r\avg{\omega}|_{1}}
 \quad\mbox{and}\quad
 \delta_\mathcal{L}=\frac{\Delta_\mathcal{L}}{-2\partial_r\avg{\mathcal{L}}|_{1}} 
\end{equation}
for the angular velocity and angular momentum profiles in TC flow 
with the total differences $\Delta_\omega=\omega_1-\omega_2$ and $\Delta_\mathcal{L}=\mathcal{L}_1-\mathcal{L}_2$.
In the low-curvature case $\eta=0.99$ analysed here, 
we also have $\delta_\omega\approx d/(2\Nus_\omega)$, 
whereas such a relation does not exist for~$\delta_\mathcal{L}$.
However, for strongly co-rotating cylinders $\delta_\omega$ overestimates the width of the boundary layer region
since then the angular velocity profiles have a significant slope in the bulk \cite{Brauckmann2013,Ostilla2013}.
As the angular momentum profile generally becomes almost flat in the bulk \cite{Brauckmann2016a}, 
the thickness $\delta_\mathcal{L}$ provides a better approximation to the size of the boundary layer region and will therefore be used here.
In figure~\ref{fig:slope-BL}
the boundary layer thickness $\delta_{\mathcal L}$ is plotted for the same data as in figure~\ref{fig:Nu-LSC}(a). It is observed that the 
boundary layer thickness of case 1 matches almost perfectly with the thermal boundary layer thickness $\delta_T$ of the RB flow at $\Ray=10^7$.  
In case 2, the differences in the thickness scales are larger; 
here, the thickness $\delta_\mathcal{L}$ is smaller than $\delta_\omega\approx\delta_T=d/(2\Nus_T)$ 
since the angular velocity profile is not flat in the central region.
As we will see in the following, these differences will affect the statistical 
properties of the TC flows and thus the agreement with RB flow.

\section{Statistical properties}
\label{sect:statistics}

\subsection{Mean vertical profiles}
\begin{figure}
	\centering
	\includegraphics[width=\textwidth]{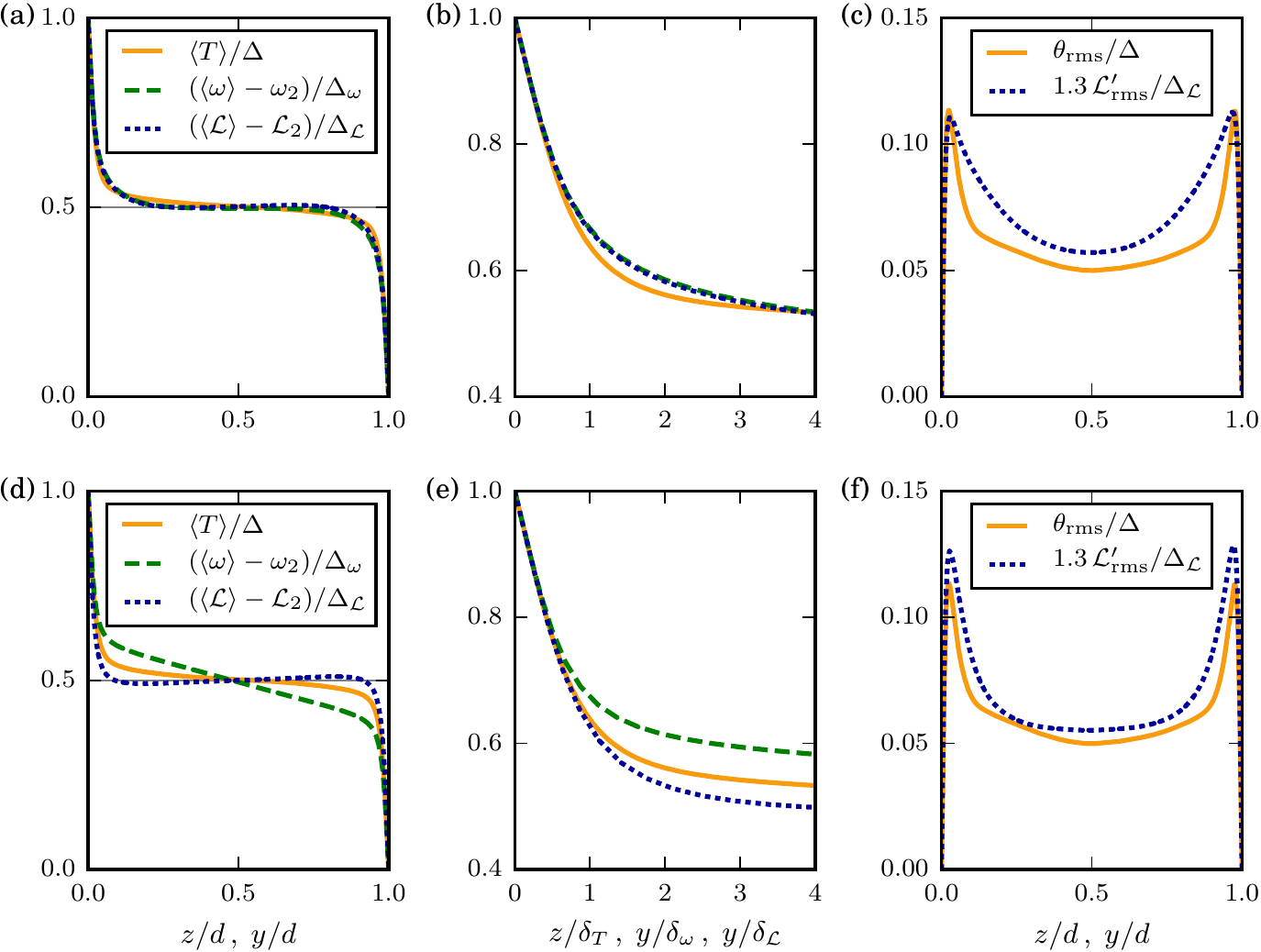}
\caption{(a) Profiles of the mean temperature $\avg{T}$ for $\Pra=0.7$ and $\Ray=10^7$  and of the mean angular velocity $\avg{\omega}$ and angular 
momentum $\avg{\mathcal{L}}$ for $\eta=0.99$, $\Rey_S=2\e{4}$ and $R_\Omega=0.023$ (case 1). 
All profiles are rescaled to the interval $[0, 1]$ using $\Delta_\omega=\omega_1-\omega_2$ and $\Delta_\mathcal{L}=\mathcal{L}_1-\mathcal{L}_2$. 
The coordinate $y=r-r_1$ is used for the TC flow. 
(b)~Magnification of the near-wall region. All coordinates are rescaled by the corresponding boundary layer thicknesses. 
(c)~Profiles of the root mean square temperature and the root mean square angular momentum. The latter is rescaled by a factor of $1.3$ for 
better comparability. 
(d,e,f) Similar comparison of the RB flow case to the TC flow run with $R_{\Omega}=0.241$ (case 2).} 
\label{fig:prof-1}
\end{figure}
In figure~\ref{fig:prof-1} we compare 
the mean profiles of temperature to the mean profiles of the angular velocity $\omega=u_{\varphi}/r$ and the angular 
momentum ${\mathcal L}=r u_{\varphi}$. The upper row displays the comparison with the TC flow at the first local maximum at 
$R_{\Omega}=0.023$ (case 1) (see figure~\ref{fig:Nu-LSC}a). The lower row repeats this comparison for the TC flow at the second local 
maximum in the $\Nus_\omega-R_{\Omega}$ relation at $R_{\Omega}=0.241$ (case 2). While the agreement with case 1 is very good, 
there are differences for case 2. Here, the angular velocity profile has a noticeable gradient in the central region, whereas the angular 
momentum is well mixed and lies closer to the temperature profile. From this comparison one can conclude that $T$ is more closely 
associated with ${\mathcal L}$  than with $\omega$. Therefore, we compare the root mean square profiles of $T$ and ${\mathcal L}$ 
fluctuations in panels (c) and (f) of figure~\ref{fig:prof-1}. The fluctuating fields are obtained by
\begin{align}
\theta({\bf x}, t)&=T({\bf x},t)-\avg{T}(z)\,, \\
{\mathcal L}^{\prime} ({\bf x}, t)&={\mathcal L}({\bf x},t)-\avg{\mathcal L}(r)\,.
\end{align}
For case 1, the peaks in $\mathcal{L}^\prime_\mathrm{rms}$ are broader than in case 2, 
which is consistent with the observation that the boundary layers are turbulent 
for case 1, but not for case 2 \cite{Brauckmann2016b}.
In the RB flow case, the boundary layer dynamics is close to laminar, and the peaks in $\theta_\mathrm{rms}$ are narrower.
Furthermore, we note that the shape of the root mean square angular velocity profile $\omega^\prime_\mathrm{rms}=\mathcal{L}^\prime_\mathrm{rms}/r^2$ (not shown here) hardly differs from $\mathcal{L}^\prime_\mathrm{rms}$  since the radius only varies by $1\%$ for $\eta=0.99$. 

\begin{figure}
	\centering
	\includegraphics[width=\textwidth]{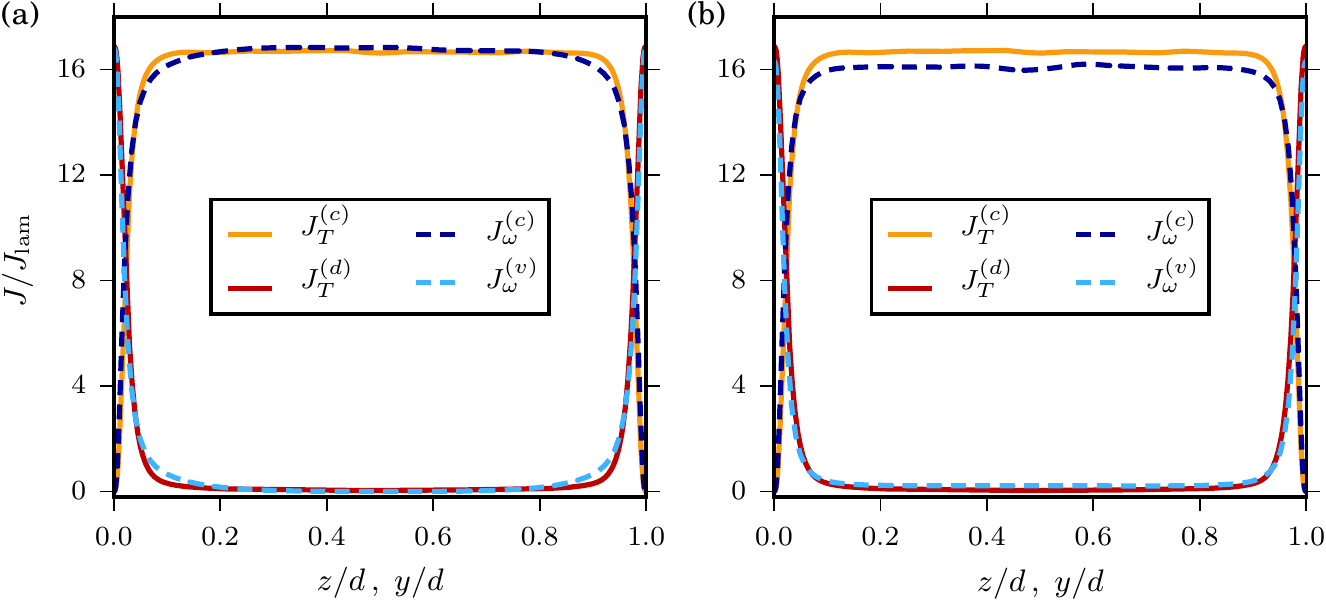}
\caption{Comparison between the contributions to the currents for (a) case 1 and (b) case 2 with the parameters described in~figure~\ref{fig:prof-1}.
The convective and diffusive currents for the heat transport in RB are shown as continuous lines, 
the corresponding convective and viscous currents for the angular momentum transport in TC as dashed lines.}
\label{fig:currents}
\end{figure}

The heat flux in RB flow is decomposed into a convective and diffusive contribution which results in
\begin{equation}
J_T=J_T^{(c)}(z)+J_T^{(d)}(z)=
\avg{u_z T}-\kappa\frac{\partial \avg{T}}{\partial z}=\Nus_T J_{T,lam}\,.
\label{eq:Nu_T-decomp}
\end{equation}
A similar decomposition into a convective and viscous contribution in the TC flow case leads to
\begin{equation}
J_\omega=J_{\omega}^{(c)}(r)+J_{\omega}^{(v)}(r)= 
 r\avg{u_r \mathcal{L}} -\nu r^3 \frac{\partial \avg{\omega}}{\partial r} 
 =\Nus_{\omega} J_{\omega,lam}\,.
\label{eq:Nu_omega-decomp}
\end{equation}
Figure~\ref{fig:currents} displays the vertical (radial) profiles. Panel (a) compares with case 1 while panel (b) compares with case 2.
As expected the dissipative contributions are significant in the boundary layers and become small in the bulk. The convective parts 
dominate the bulk and drop to zero at the walls due to the no-slip boundary conditions. 
Furthermore, the sum of both transport contributions remains constant across the whole layer in both systems.
It can be seen again that profiles of case 1 show better agreement with RB flow than the profiles of case 2. 
Since in the latter case the angular velocity profile is not flat in the central region (cf. figure~\ref{fig:prof-1}d), 
the viscous contribution $J_{\omega}^{(v)}$ is larger than the corresponding diffusive part $J_T^{(d)}$, 
which additionally results in a smaller convective contribution $J_{\omega}^{(c)}$.

\subsection{Probability density functions}
\begin{figure}
	\centering
	\includegraphics[width=\textwidth]{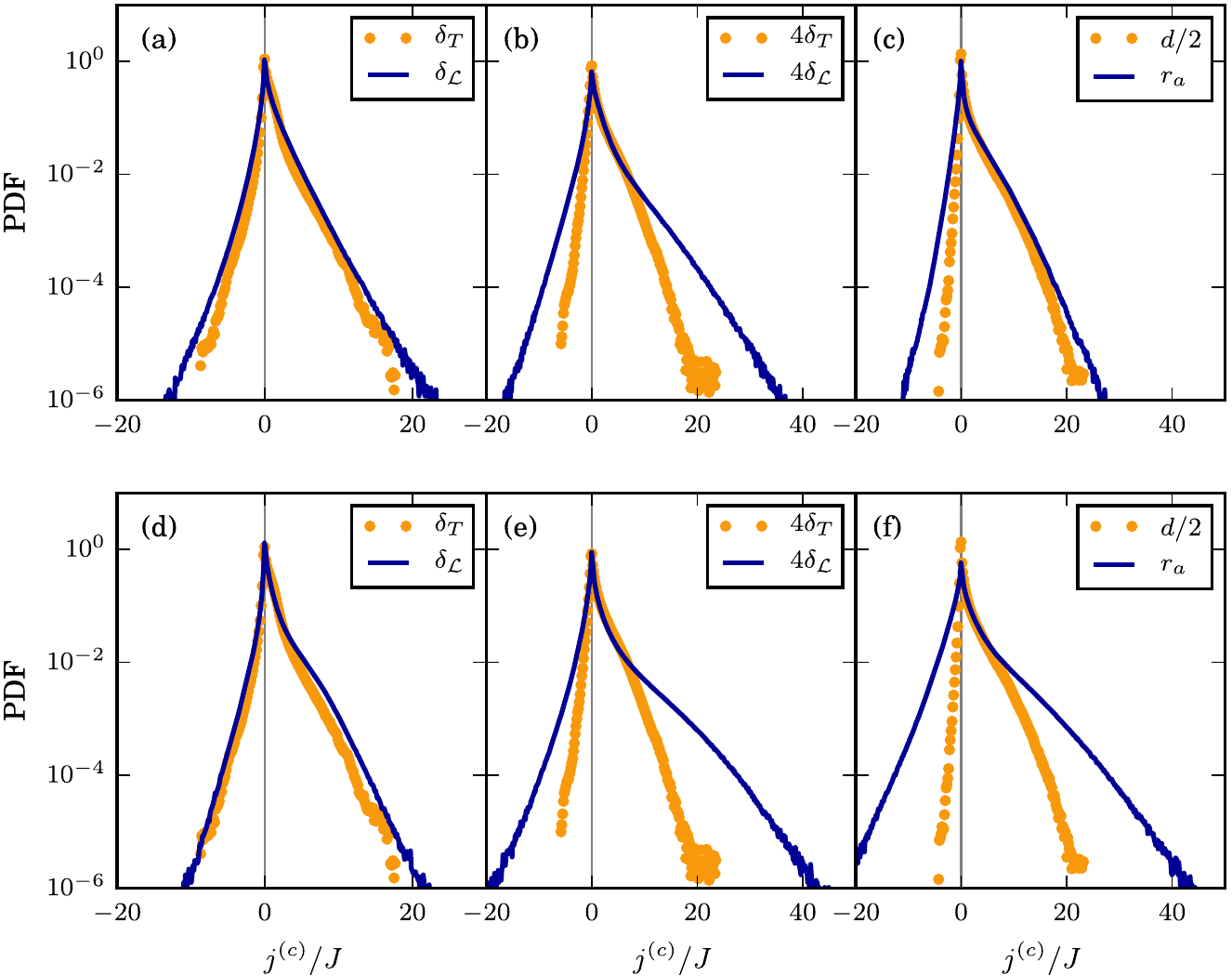}
\caption{Probability density functions (PDFs) of the local convective currents $j_T^{(c)}= u_z \theta$ (filled circles) 
and $j_\omega^{(c)}=ru_r\mathcal{L}'$ (solid lines) computed at the boundary layer thickness $\delta_T$ ($\delta_\mathcal{L}$) in panels (a, d), 
at $4\delta_T$ ($4\delta_\mathcal{L}$) in panels (b, e) and at the centre $d/2$ ($r_a$) in panels (c, f). 
The local currents are normalized by the corresponding total mean currents $J_T$ and $J_\omega$.
Panels (a, b, c) are for case 1 and panels (d, e, f) are for case 2 for the same parameters 
as in~figure~\ref{fig:prof-1}. Here $\theta=T-\avg{T}$ and $\mathcal{L}'=\mathcal{L}-\avg{\mathcal{L}}$ denote the fluctuations around the corresponding 
time- and area-averaged profiles.}
\label{fig:jpdf}
\end{figure}

We now refine the analysis and report the statistics of the convective currents in averaging surfaces at different 
distance from the inner (bottom) wall. First, it is important to note that only the temperature and angular momentum 
fluctuations $\theta$ and $\mathcal{L}^\prime$ that deviate from the corresponding mean profile contribute to the net 
transport through the averaging surfaces, since 
$\langle u_z\rangle=\langle u_r\rangle=0$ by incompressibility, and therefore
$\avg{u_z T}=\avg{u_z \theta}$ and $r\avg{u_r \mathcal{L}}=r\avg{u_r \mathcal{L}^\prime}$.
Therefore, we study
the local convective currents based on the fluctuations $\theta$ and $\mathcal{L}^\prime$ instead of the total fields.
Figure~\ref{fig:jpdf} compares the probability density functions (PDFs) of the local convective angular momentum current $j_{\omega}^{(c)}=
ru_r\mathcal{L}'$ for cases 1 (top row) and 2 (bottom row) with the local convective heat current $j_T^{(c)}=u_z\theta$ for planes at 
different distances from the wall.  All quantities are normalized by the corresponding mean currents $J_\omega$ and $J_T$.
In all cases, it is observed that the skewness of the distributions increases with the distance of the analysis surface from the inner 
(bottom) wall. The net convective transport has to be positive, and its share of the total transport increases towards the bulk. 
It is also observed that the tails of the PDFs of TC flow for case 2 deviate strongly from the ones for RB flow
away from the boundary layer.

The trend is different for the comparison of case 1 with
RB flow. While the largest 
differences arise for the data at $4 \delta_\mathcal{L}$, the agreement is very good for the data taken at $\delta_\mathcal{L}$ and $r_a$.  
The region just above the boundary layer thickness is dominated by rising plumes and recirculations next to the plumes. 
It is sometimes also denoted as the plume mixing layer \cite{Castaing1989}. 
The reason for the differences in the width of the tails in panels (b) and (e) of figure~\ref{fig:jpdf} could 
therefore be related to the shape of the plumes and the frequency of their detachment, 
which differ between RB and TC flow as will be shown in the next subsection.

The observation that the local fluctuations in case 2 are enhanced compared to case 1 
can be understood by analysing the components that form $j_{\omega}^{(c)}=ru_r\mathcal{L}'$. 
Since the fluctuation amplitude of $\mathcal{L}'$ varies little between both cases (cf.~figure~\ref{fig:prof-1}(c,f)), 
and the radius $r$ remains unchanged, the difference must occur in the radial velocity $u_r$.
In \cite{Brauckmann2016a} it was shown that the fluctuation amplitude $(u_r)_\mathrm{rms}$ varies with mean rotation ($R_\Omega$)
and in case 2 it is twice as large as  in case 1.
This increase is partly caused by a strengthening of the mean Taylor vortices, cf.~figure~\ref{fig:Nu-LSC}(b).
The stronger $u_r$ fluctuations result in wider tails in figure~\ref{fig:jpdf}(e,f) 
but do not significantly affect the distribution at $\delta_\mathcal{L}$ 
where the radial velocity is restricted due to the proximity of the cylinder wall.

\begin{figure}
	\centering
	\includegraphics[width=\textwidth]{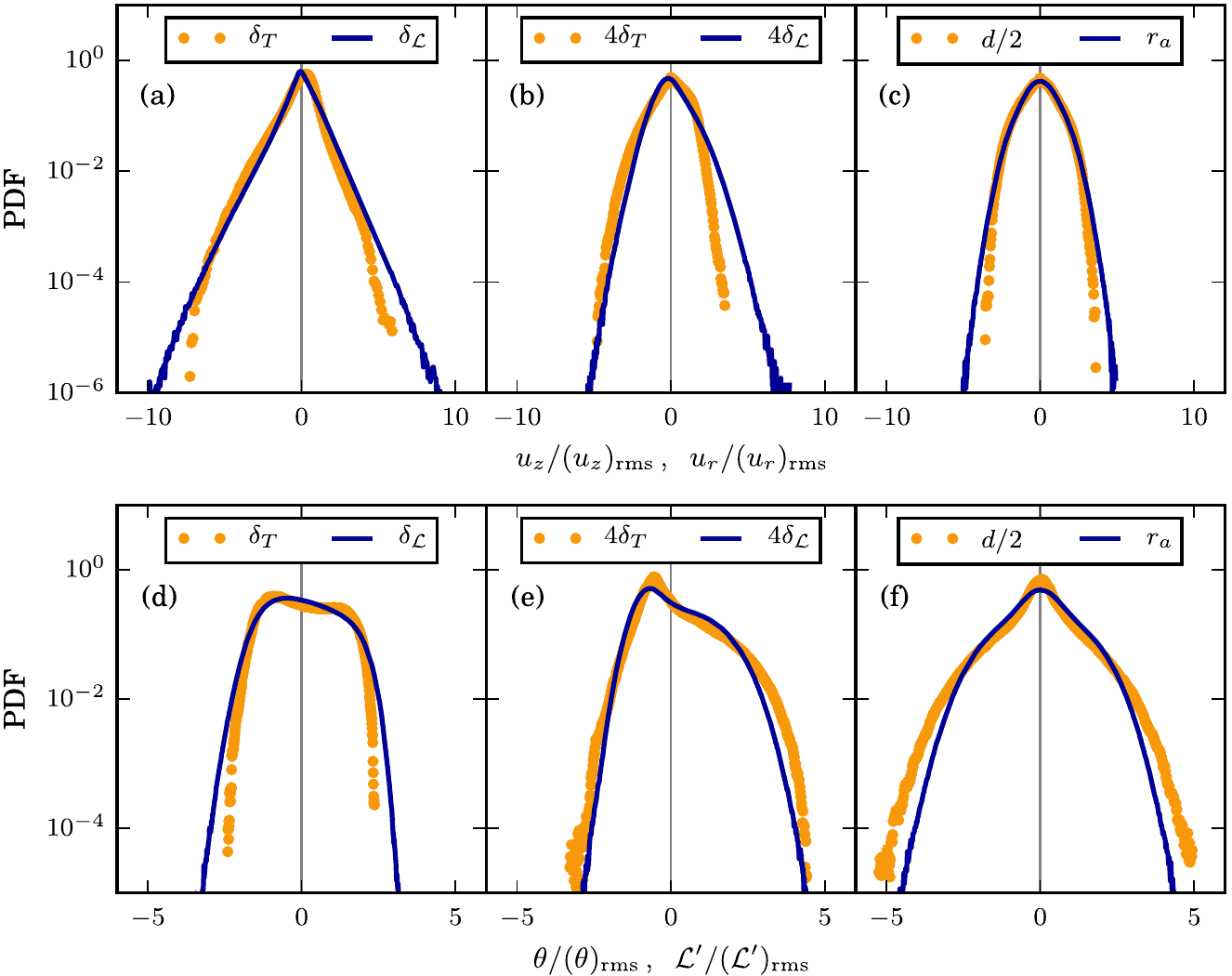}
\caption{Comparison of the PDFs of individual components of the turbulent fields. 
All data are for case 1 and the RB flow. 
In the upper row, the PDFs of the axial velocity $u_z$ (RB, filled circles) and radial velocity $u_r$ (TC, solid lines) are compared,
in the lower row the PDFs of the temperature $\theta$ (RB, filled circles) and angular momentum $\mathcal{L}^\prime$ (TC, solid lines). 
Data are obtained for $\delta_T$ ($\delta_\mathcal{L}$) in panels (a, d), 
at $4\delta_T$ ($4\delta_\mathcal{L}$) in panels (b, e) 
and at the centre $d/2$ ($r_a$) in panels (c, f). 
Each PDF is normalized by its corresponding root mean square value.}
\label{fig:jdpdf}
\end{figure}

In figure~\ref{fig:jdpdf} we compare individual components 
of  the transport currents of angular momentum and heat. They are $u_z$ and $\theta$
in the RB case, and the radial velocity $u_r$ and angular momentum $\mathcal{L}^\prime$ in TC flow. 
It can be observed that the agreement between case 1 
and the RB flow is good. For case 2 (not shown here),
the deviations  of the individual PDFs of $\theta$ and $\mathcal{L}^\prime$ were larger.
In a turbulent flow one expects Gaussian statistics for the individual components of the 
velocity field. In figure \ref{fig:jdpdf}(a) exponential tails are observed for the PDFs of $u_z$ and $u_r$ at the height of (thermal) boundary 
layer thickness. This is a clear statistical fingerprint for an enhanced intermittency in the near-wall region which is connected with the plume 
formation. Also, in figure \ref{fig:jdpdf}(b) a fatter tail for the radial component is detected which confirms our observation in figure \ref{fig:jpdf}(b). 

The distributions of the temperature and angular momentum fluctuations (see lower row of figure \ref{fig:jdpdf}) are skewed and take a symmetric 
shape in the midplane only as shown in figure \ref{fig:jdpdf}(f). For all three distances the tails of both  PDFs are in very good agreement. The distribution 
in the midplane is again not Gaussian which has been reported already in \cite{Emran2008}. The specific cusp-like form around the origin and the increasingly
pronounced exponential tails in the PDF of temperature fluctuations have been discussed, for example, by Yakhot on the basis of a hierarchy
of momentum equations for the temperature fluctuations \cite{Yakhot1989}. Interestingly, even such specific 
details of the small-scale statistics prevail in our comparison between RB and TC flow.
\begin{figure}
	\centering
	\includegraphics[width=1.0\textwidth]{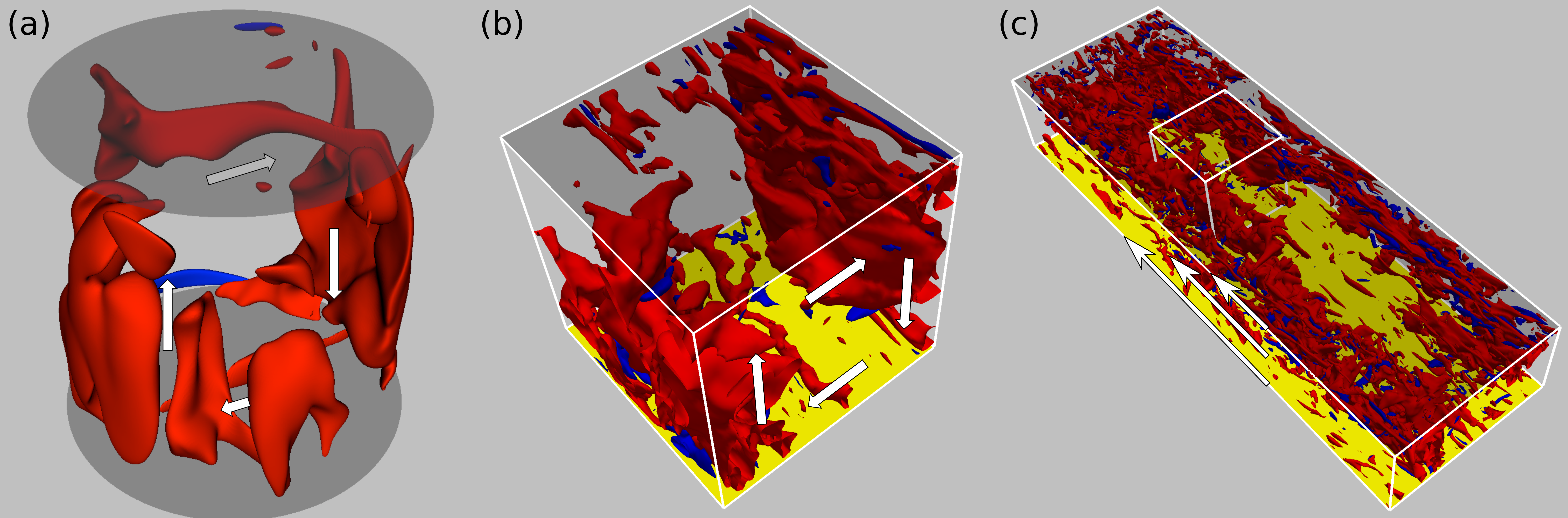}
\caption{Isosurface snapshots of the convective transport currents $j_T^{(c)}$ for RB flow and $j_\omega^{(c)}$ for TC flow (case 1). 
Isolevels $j^{(c)}/J=-3$ (blue) and $3$ (red) are shown.
(a)~RB flow with top and bottom walls indicated by semi-transparent planes. 
(b,c)~TC flow with the inner and outer cylinder indicated by a yellow and semi-transparent surface, respectively. 
(b)~Magnification of the segment marked in (c) with an extension of $1d$ in each direction, to match the aspect
ratio of the RB flow domain. 
(c)~Simulated fraction of the annular domain. 
The arrows indicate the large-scale circulation in (a), the Taylor vortex in (b) and the velocity profile in (c).}
\label{fig:isosurf}
\end{figure}

\subsection{Relation between transport and flow structures}    
In the last subsections, we identified small differences in the statistical properties of the RB flow and 
case 1 of TC flow and 
attributed them to differences in the flow structures. 
The spatial organization of currents
is shown in figure~\ref{fig:isosurf}, where isosurfaces of the convective current (the top row of figure~\ref{fig:jpdf})
for the levels $j^{(c)}/J=\pm 3$ are plotted.
Panels (a) and (c) show the full simulation domains in both cases.
The magnification in figure~\ref{fig:isosurf}(b) displays a section of the TC flow with the same aspect
ratio as the RB domain in panel (a). 
Since the net transport current is positive, on average, there is a larger volume fraction of red isosurfaces than blue ones. 
The large coherent regions of high convective current, which occur near the sidewall in panel (a) and near the left and right 
surfaces in panel (b), coincide with the upward and downward motion of the large-scale circulation in RB and a Taylor vortex in TC
(see white arrows in panels (a,b)). Consequently, the positive tails in 
figure~\ref{fig:jpdf}(c) are related to this large-scale motion. 
The similarity in the large-scale organization of the currents explains why the differences in geometry (cylindrical vs. rectangular domain) have a minor influence on the statistical properties in the middle, cf.~figure~\ref{fig:jpdf}(c).
The isosurfaces of the TC flow are more fragmented and less smooth
than for the RB flow, which indicates a higher level of fluid turbulence in the system. Quantitatively, we find in RB flow for
the large--scale Reynolds number $\Rey_{rms}= u_{rms} d/\nu=675$, with the root mean square velocity $u_{rms}$ 
calculated from all three velocity components in the entire cell \cite{Scheel2014}. 
This is significantly smaller than the corresponding Reynolds numbers  in TC flow (with the mean rotation subtracted),
which are $\Rey_{rms}=2251$ and $\Rey_{rms}=2712$ for the cases 1 and 2, respectively.
It can thus be expected that the boundary
layers in the RB case are still close to laminar, while the ones in case 1 of TC flow are already turbulent \cite{Brauckmann2016b}. 
Specifically, we find for the boundary-layer Reynolds numbers, 
defined based on the boundary layer thickness and shear across the boundary layer, 
values of $\sim30$ for RB flow \cite{Scheel2014} and of $\sim300$ and $\sim200$ for the TC cases 1 and 2, respectively \cite{Brauckmann2016b}.
The turbulent fluctuations in the TC boundary layer account for the deviations in the tails of the PDFs and 
the slight deviations in the area-averaged profiles, in particular at the heights of $z=4 \delta_T$ and $y=4 \delta_{\cal L}$,
respectively, in figure~\ref{fig:jpdf}(b).

\section{Conclusions}
\label{sect:conclusions}

In the present work we discussed a direct comparison of the statistical properties of Rayleigh--B\'{e}nard (RB) convection and 
Taylor--Couette (TC) flow. The comparison is motivated by analogies of dimensionless system parameters (such as Rayleigh 
and Taylor numbers), the same form of the energy balances, (\ref{eps_temp}) and (\ref{eps_omega}), and the similarities in the currents of 
heat and angular momentum
(see also references \cite{Eckhardt2007,Eckhardt2007a,Bradshaw1969,Dubrulle2002}). 

Our study shows that the operating point for a specific comparison between TC and RB flows 
can be determined
by choosing corresponding values of Nusselt numbers since the Nusselt number defines the 
boundary layer thickness and hence the transport properties. We also find that a better characterization 
of TC flow can be based on the pair of
shear Reynolds and rotation numbers, $(\Rey_S, R_{\Omega})$, than on Taylor and quasi-Prandtl 
numbers, $(\Tay, \sigma)$,
since the latter do not reflect
the mean rotation of the cylinders.
We demonstrated that for sufficiently large shear Reynolds number $\Rey_S$, multiple TC flow cases at different rotation numbers can 
have the same Nusselt number as RB convection, i.e. the same amount of angular momentum is transported between the cylinders 
in TC flow as heat from the bottom to the top in the RB case. The comparison also shows that the case with the smaller rotation number 
$R_\Omega$ (case 1) provides a better agreement with RB flow than the case of larger rotation number. 
For this pair of flows, a remarkable agreement between
mean profiles as well as probability density functions of fluctuating quantities is found. 

Studies of the mean profiles and the PDFs of the convective currents show that the differences between RB flow and TC flow case 1 are 
most pronounced in the mixing layer above the (thermal) boundary layer.
They can be attributed to the strong fluctuations in this 
region which are connected with the detachment of plumes and other differences in the dynamics: the boundary layers in the convection case are still very close 
to being laminar, but in the TC system they are already turbulent. The differences should, therefore, become smaller when the boundary layers 
in RB become turbulent as well.

The TC flow case 2, which is characterized by a larger mean rotation ($R_{\Omega}$), shows greater differences to the RB case. 
As a consequence of rotation, the angular velocity profile has a significant gradient in the central region, which results in a higher (lower) 
dissipative (convective) transport current than in the RB case. Furthermore, enhanced radial velocity fluctuations and stronger mean 
Taylor vortices occur for case 2 and lead to broader PDFs of the convective current away from the boundary layer, which differ from 
the heat flux distributions in RB flow. This demonstrates that the mean rotation determines how well the transport characteristics of 
TC and RB flow are comparable.

The comparison presented here shows that for judiciously chosen pairs of parameters in RB and TC flow one
can actually relate their transport properties in detail, both in the mean and in the fluctuations, thereby confirming
the analogies between the \emph{twins of turbulence} \cite{Busse2012} for a larger set of properties.

\acknowledgements
We would like to thank M. S. Emran and R. du Puits for scientific discussions at the beginning of this work. 
JS~acknowledges computational resources 
provided by the John von Neumann Institute for Computing within Supercomputing Grant HIL09.  HB and BE thank 
M. Avila for providing the code used for the TC simulations and acknowledge computational resources at the LOEWE-CSC in Frankfurt.
The paper was written during a workshop at the Lake Arrowhead Conference Center, and BE and JS would like to thank the 
Institute of Pure and Applied Mathematics (IPAM) of the University of California Los Angeles for financial support.

\bibliographystyle{plain}

\bibliography{library.bib}

\end{document}